\documentclass[useAMS,usenatbib,usegraphicx]{mn2e}

\usepackage{fixltx2e}

\def\apss{Ap\& SS}
\def\apj{ApJ}
\def\apjl{ApJL}
\def\aap{A\& A}

\def\araa{ARA\&A}
\def\mnras{MNRAS}

\def\aj{{AJ}}
\def\apjs{ApJS}

\def\mnras{{MNRAS}}
\def\pasj{{PASJ}}

\def\'#1{\ifx#1i{\accent"13\i}\else{\accent"13#1}\fi}
\def\alamenos#1{$^{-#1}$}
\def\ala#1{$^{#1}$}

\newcommand{\grad}{^{\mbox{\small o}}}

\def\kms{km sec\alamenos 1}

\def\be{\begin{equation}}
\def\ee{\end{equation}}

\def\Eg{{E_{\rm grav}}}

\def\Wext{W_{\rm ext}}

\def\Msun{M_\odot}

\newcommand{\pcc}{\mbox{\,cm}^{-3}}
\newcommand{\pc}{\mbox{\,pc}}

 \title[Tidal foces as a regulator of star formation in Taurus ] {Tidal forces as a regulator of star formation in Taurus  }

 \author[Ballesteros-Paredes, et al.]
{ \parbox{7.0in}{
Javier Ballesteros-Paredes\ala 1,
Gilberto C.\ G\'omez\ala 1, 
Laurent Loinard\ala 1,
Rosa M.\ Torres\ala 1, 
and  B\'arbara Pichardo\ala 2 \\
} \\ \ala 1 Centro de Radioastronom\'ia y Astrof\'isica,
            Universidad Nacional Aut\'onoma de M\'exico, \\
            Apdo. Postal 72-3 (Xangari), Morelia,
            Michoc\'an 58089, M\'exico \\
            j.ballesteros,g.gomez,l.loinard,r.torres@crya.unam.mx\\
     \ala 2 Instituto de Astronom\'ia,
            Universidad Nacional Aut\'onoma de M\'exico, \\
            Apdo. Postal 70-264, 04510, M\'exico, D.F., M\'exico\\
            barbara@astroscu.unam.mx
}

\begin{document}

\date{Submitted to MNRAS, \today}

\pagerange{\pageref{firstpage}--\pageref{lastpage}} \pubyear{2006}

\maketitle

\label{firstpage}

\begin{abstract}

  Only a few molecular clouds in the Solar Neighborhood exhibit the
  formation of only low-mass stars.  Traditionally, these clouds have been
  assumed to be supported against more vigorous collapse by magnetic
  fields. The existence of strong magnetic fields in molecular clouds,
  however, poses serious problems for the formation of stars and of
  the clouds themselves. In this {\em Letter}, we review the
  three-dimensional structure and kinematics of Taurus --the archetype
  of a region forming only low-mass stars-- as well as its orientation
  within the Milky way. We conclude that the particularly low
  star-formation efficiency in Taurus may naturally be explained by
  tidal forces from the Galaxy, with no need for magnetic regulation
  or stellar feedback.

\end{abstract}



\section{Introduction}\label{sec:intro}

Few nearby molecular clouds (e.g.\ Taurus, Chamaeleon I and II) are
observed to form only low-mass stars. Assuming a universal initial
mass function, the lack of high-mass young stars in these regions
indicates a low overall star-forming rate compared with clouds like
Orion where both low- and high-mass stars are actively being formed.
Traditionally, the clouds forming only low-mass stars have been
assumed to be supported against more vigorous collapse by magnetic
fields. In such a scheme, stars can form only after substantial
magnetic flux has been removed locally via ambipolar diffusion
\citep[e.g.,][]{SAL87}.  The Taurus Molecular Cloud (TMC) is often
cited as the archetype for this picture of isolated, low-mass star
formation.

There are, however, some difficulties with this scenario. On the one
hand, strong magnetic fields ought to prevent the formation of
molecular clouds by large-scale compressions in the first place
\citep{HBB01}.  In addition, there is a problem with the
synchronization of star formation along large distances. While
molecular clouds have dynamical timescales of the order of 10--20 Myr,
most active star-forming regions (i.e.\ those still containing
molecular gas) have populations with ages $\le$ 3~Myr
\citep{HBB01}. The lack of old stars associated with molecular clouds
\citep[the so-called post-T Tauri problem, see][]{Herbig78,
Herbig_etal86} has been explained in terms of rapid assembling of
molecular clouds by large-scale flows, which may be able to trigger
star formation over large regions almost simultaneously. Such a rapid
assembling of molecular clouds and synchronized events of star
formation over large distances require magnetic fields to {\em not} be
dominant \citep{HBB01}.
{ Moreover, the ambipolar diffusion timescale is not unique.  It
depends, among other parameters, on the ionization fraction, which in
turn depends on the precise local shielding conditions. Differences in
the degree of ionization and magnetic field intensities should
introduce an unobserved spread of at least several Myr in the onset of
star formation. This brings us back to the post-T Tauri problem
\citep{BH07}.  }

Different numerical work has examined the picture of rapid molecular
cloud assembling from different points of view and found it to be a
viable mechanism \citep[see][ and references therein]{BKMV07}.  An
important difficulty, however, is the low star formation efficiency
observed in actual molecular clouds, compared with those reported in
simulations.  As discussed by \cite{Heitsch_Hartmann08}, most
simulations are performed in closed boxes, with no stellar energy
feedback. In such a situation, the amount of mass in collapsed objects
after one crossing time is usually large, with values between 10\% and
30\%, depending on the mass and the level of turbulence of the model
\citep{KHM00, VBK03, VKB05}.  In comparison, typical values of the
star formation efficiency observed in molecular clouds is only a few
percent \citep{Myers_etal86}. When feedback from massive stars is
included in the simulations, the measured efficiencies are
significantly smaller \citep{PVP95}, suggesting that massive stars are
a key ingredient in regulating the efficiency of star formation
\citep[see also ][and references therein]{BP04P}.

{ In regions where no massive stars are formed, however, a
different mechanism is clearly required. And since magnetic regulation
brings a number of additional problems, it is worth looking for
alternative possibilities. } 
Recently, \citet[][= Paper I]{BP_etal09}
have analyzed the complete gravitational content of molecular clouds
within a given spiral galaxy.  They write the total gravitational
energy of a molecular cloud, $W$, as follows:

\begin{equation}
  W = - {1\over 2} \int_V \rho \Phi_{\rm cl} dV - \int_V x_i\ \rho\
 {\partial \Phi_{\rm ext} \over \partial x_i} \ dV,
\label{eq:Wtot}
\end{equation}
where $\rho$ is the density, $\Phi_{\rm cl}$ is the gravitational
potential due to the mass of the cloud, i.e., the mass inside its
volume $V$, $\Phi_{\rm ext}$ is the gravitational potential due to the
mass outside the cloud, and $x_i$ is the $i$-th component of the
position vector. The first term on the right is the gravitational
energy $\Eg$, while the second is the energy due to the mass outside
of the cloud, which we call the tidal energy. This second term may
cause a compression or a disruption of molecular clouds, depending on
their size, position and orientation within the host galaxy. Thus,
tidal interactions may play a significant role in the overall
stability of molecular clouds, and, therefore, on the efficiency of
star formation within them.

In this {\em Letter}, we investigate the role that tidal interactions
might play in the regulation of star formation in the TMC. In
\S\ref{sec:model}, we examine the three-dimensional structure and the
orientation of Taurus within the Milky Way using recent data. We then
calculate the relative contribution of self-gravity and tidal
interactions for such a configuration (\S\ref{sec:results}) and
discuss our results in \S\ref{sec:discuss}. The conclusions are given
in \S\ref{sec:conclusions}.

\begin{figure}
\includegraphics[width=0.95\hsize]{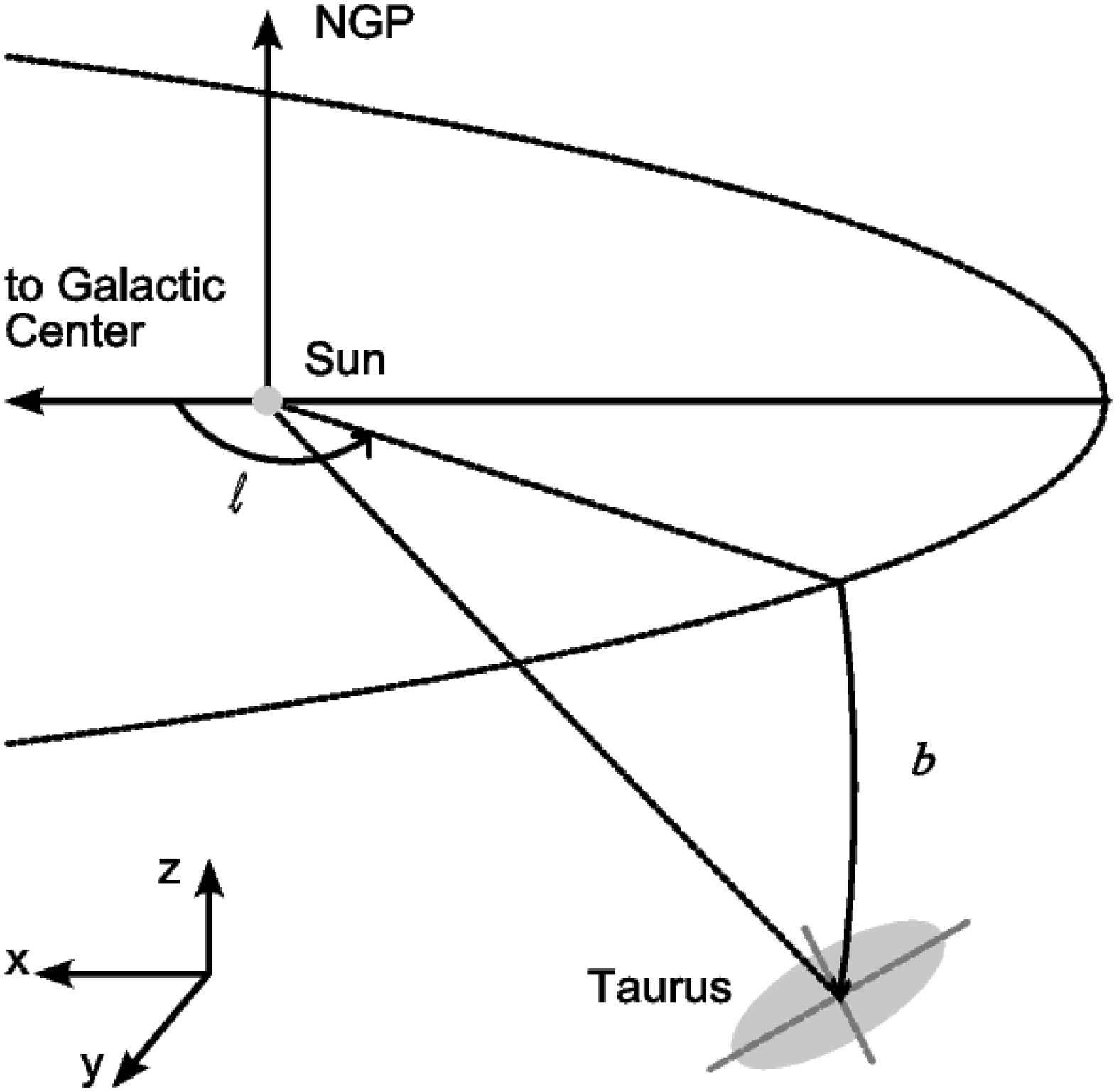}
\caption{Schematic view of the TMC, according to the distances and
  positions reported by \citet{Loinard_etal05, Loinard_etal07, Torres_etal07,
    Torres_etal09}.}
  \label{fig:tmc}
\end{figure}

\section{Three dimensional structure of Taurus}\label{sec:model}  

In CO maps, the TMC extends for about 10 degrees on the sky, with
filaments that have aspect ratios between 5:1 and 10:1
\citep[e.g.,][]{Goldsmith_etal08}. It has a total molecular mass
between 10\ala 4 and $\sim$ 2.4$\times 10^4\Msun$
\citep{Goldsmith_etal08}, and is located roughly towards the Galactic
anti-center (at $l$ $\sim$ 170$^\circ$--176$^\circ$) but at a Galactic
latitude of --15$^\circ$ to --{16$^\circ$}.  Recent multi-epoch VLBA
observations have provided distances to several young stars located
across the TMC complex with accuracies better than 1\% (corresponding
to absolute errors of $\sim$ 0.5--1 pc) \citep{Loinard_etal05,
Loinard_etal07, Torres_etal07, Torres_etal09}. The obtained distances
range from about 160~pc for HP Tau, near the eastern edge of the
complex at $(l,b) \sim ( 175\grad, -16\grad )$ \citep{Torres_etal09},
down to about ~130~pc for the closest stars, Hubble 4 and HDE 283572,
in the western part of the TMC at $(l,b) \sim (170\grad, -15\grad)$
\citep{Torres_etal07}. This situation is quite unlike that in the core
of Ophiuchus \citep{Loinard_etal08} or the Orion nebula
\citep{Menten_etal07} where different stars are found at very similar
distances.

The properties mentioned above indicate that it is appropriate to
model the TMC as a 10$^4$ $M_\odot$ elongated filament (a prolate
spheroid) centered at a distance of $145\pc$ from the Sun in the
direction $(l,b) = (172.5\grad, -15\grad)$ (see Fig.~\ref{fig:tmc} for
a schematic view).  This places the TMC about $37.5\pc$ below the
Galactic plane. The long axis of the spheroid was taken to be of
$32.37\pc$ long and, assuming an angular width for the cloud of about
$5\grad$, the short axes are $5\pc$ in length. The density of the
spheroid was taken to be constant; for our choice of parameters, its
value is $n=405\pcc$.

We performed an energetics analysis similar to that presented in Paper
I, but with one important difference. In our previous work, every
parcel of the test cloud had a velocity given by the circular
velocity.  
{ As most nearby star forming cloud, however, the TMC
has a substantial peculiar velocity \citep{Torres_etal09}}.
Therefore, although the calculation of the tidal energy $\Wext$ is
performed in the standard of rest of the center of the spheroid, the
effective potential must involve the peculiar velocities of the
filament.  In order to account for those velocities, we calculated the
components of the peculiar velocities $(u,v,w)$ using radial
velocities from the CO observations \citep[][see also Figs. 2 and 3 in
Ballesteros-Paredes et al.  1999]{Ungerechts_Thaddeus87}, and the
proper motions of the stars reported by
\citet{Torres_etal07,Torres_etal09}. We note that the proper motions
have been determined with a very good accuracy ($\pm$ 0.15 mas
yr\alamenos 1 in the worst case). However, since the line profile of
the gas has some spread around the maximum intensity, we have used
radial velocities (express relative to the Local Standard of Rest)
ranging from 5 to 5.5~\kms near the eastern edge of the cloud, and
from 5.5 to 6~\kms for the western part.

\section{Results}\label{sec:results}

\begin{figure}
\includegraphics[width=0.95\hsize]{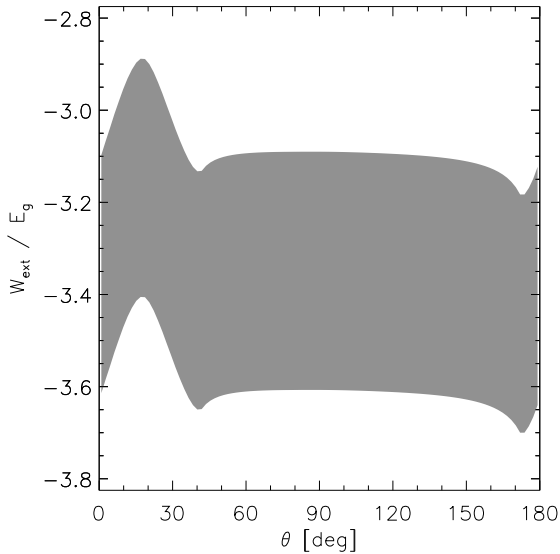}
\caption{$\Wext/\Eg$ ratio for an elongated spheroid at the position
  and orientation, and with the kinematical properties of the TMC as a
  function of galactocentric angle $\theta$. 
  {The shadded area covers the range of radial velocities at the
  extrema of the spheroid (see \S\ref{sec:model}) }.
Since the potential is bisymmetric, we plot the results only for the
  range $\theta ~ \in [0,180]$~degrees.  }
  \label{fig:Wext}
\end{figure}

As in Paper I, the gravitational potential used to calculate the tidal
energy $\Wext$ includes a Galactic axisymmetric background potential
that represents a bulge, a flattened disk, and a massive halo, and a
bisymmetric potential describing a logarithmic spiral pattern. Our
choice of parameters describing this potential reproduces, in
particular, the Oort constants, the rotation curve and the local
escape velocity \citep[see][and references therein]{Pichardo_etal03}.
The exact position of the TMC with respect to the Galactic stellar
spiral arms is not well known.  Thus, we have calculated the ratio
between the tidal and the gravitational energy, $\Wext/\Eg$, as a
function of galactocentric angle, $\theta$ (Fig.~\ref{fig:Wext}). The
different curves are the results of our calculations assuming a
slightly different radial velocities, in order to account for the
scatter in the CO emission (see \S\ref{sec:model}).  Since the ratio
$\Wext/\Eg$ scales with density as $1/\rho$ (see Paper I for details),
it is straightforward to scale it to a different total mass.
 
Fig.\ \ref{fig:Wext} shows some important points. First of all, the
ratio $\Wext/\Eg$ is always negative, with values between --3 and
--3.6 (shaded region), depending on the detailed velocity field
assumed for the cloud.  This indicates that the tidal energy $\Wext$
acts against the gravitational energy, i.e., by
trying to disrupt the cloud.  Second, for this configuration, the
tidal energy is larger than the gravitational energy for any azimuthal
angle.  Third, even close to the spiral arms ($\theta \sim 15^\circ$
in our figure), an elongated cloud highly aligned with the
galactocentric radius will also be disrupted.

We have performed the same calculations for smaller clumps at the same
position, but with different aspect ratios, densities, and sizes.  Our
results indicate that for smaller and denser regions, the situation is
reversed and the gravitational energy exceeds the tidal energy by
factors of 10 to 10\ala 5, depending on the properties of the
clumps/cores.  The situation considered in these calculations
correspond to individual clumps within the TMC, such as Heiles Cloud
2, or Lynds 1495 \citep[see][for details]{Goldsmith_etal08}, or to
individual dense cores like TMC-1C or Lynds 1517.
{This means that if the volume filling factor of the gas is
  smaller than unity, the small, compact, dense fragments will
  collapse, but the cloud as a whole will not. By disrupting the
  cloud, tidal forces prevent global collapse.}

\section{Discussion}\label{sec:discuss}

As mentioned in \S \ref{sec:intro}, magnetic fields and massive stars
have been the usual mechanisms invoked to explain the low efficiency
of star formation observed in star-forming regions. In the case of
Taurus, feedback by massive stars clearly cannot be invoked since
there are no such stars. Large magnetic fields, on the other hand, are
difficult to reconcile with the synchronized star formation observed
in molecular clouds in the Solar Neighborhood.

About ten years ago, it was proposed that molecular clouds in general,
and Taurus in particular, could be produced by converging large-scale
flows \citep{BHV99}, explaining how star formation can occur
simultaneously in dynamically disconnected regions.  \citet{HBB01}
pointed out that the interstellar gas in the Solar Neighborhood
becomes gravitationally unstable at the same time that it becomes
molecular, and that typical magnetic fields are not strong enough to
inhibit rapid molecular cloud and star formation.  More recently,
different authors have reported that turbulent motions may have a
gravitational origin \citep{BH04, BP06, VS_etal07, HB07, Field_etal08,
Heitsch_Hartmann08}. This revived the idea originally proposed by
\citet{Goldreich_Kwan74} that the supersonic linewidths have a
gravitational origin.  In particular, \citet{VS_etal07} showed that
collapsing clouds develop a ``virial'' type relationship\footnote{It
has been proposed to call it ``energy equipartition'' relationship,
rather than virial relationship, since observations do not allow us to
distinguish between energy equipartition and true virial balance
\citep{BP06}.}, in which kinetic and gravitational energy are within a
factor of two of each other.  Moreover, \citet{Heitsch_Hartmann08}
have found that, although molecular clouds and their substructures are
formed by colliding turbulent flows, some degree of gravitational
contraction must occur along the direction perpendicular to the
collision of the streams, to allow molecular cloud and star formation.

All the work mentioned above suggests that molecular clouds must be, to
some degree, in a state of global collapse that typical magnetic
fields cannot detain. In this situation, magnetic support cannot be
invoked to regulate star formation, and massive stars are seemingly
the only agents able to keep the star formation efficiency at a
reasonably low value. But what regulates star formation in clouds,
like Taurus, which do not harbor massive stars?

The results of the previous section allow us to propose a solution.
Like every known large molecular cloud, the TMC has a mass much larger
than its Jeans mass, so it could be collapsing vigorously. Because of
it position and orientation within the Milky Way, however, it appears
to suffer significant large-scale tidal disruptions. This ought to
prevent global collapse and limit the efficiency of star-formation.
Note, however, that tidal disruption are irrelevant at small scales
(see \S \ref{sec:results}), so lateral collapse
\citep{Heitsch_Hartmann08} and star formation can proceed in dense
cores (as is indeed observed).

According to our results, what defines whether a given molecular cloud
develop a large or a small star formation rate is its particular
position and orientation in the Galaxy. It would be interesting to
test this proposal with other regions of star-formation. According to
\citet{HB07}, Orion has a large degree of global collapse. Our results
would then suggest that this complex should not be very elongated
along the galactocentric radius. Existing observations in the region
of the Orion Nebula \citep{Menten_etal07} are consistent with this
idea, but the distance to young stars spread over a more extended area
will have to be measured to test our prediction. Another region where
this proposal could be tested is Ophiuchus, where low- and
intermediate mass stars are present. Although precise distances have
been measured for this region \citep{Loinard_etal08}, the
three-dimensional structure is still not well constrained because of
possible confusion with background stars \citep{Loinard_shanghai}.
{ Perseus would also be an interesting region to study because,
like Taurus, it appears to have a filamentary structure. Again,
however, an accurate distance is only known for one part of the
complex \citep[NGC~1333][]{Hirota_etal08}}.

\section{Conclusions}\label{sec:conclusions}

Our analysis of the full gravitational content of the Taurus Molecular
Cloud, i.e., considering not only the gravitational energy, but also
the tidal contribution from the Galaxy, indicates that TMC must be
suffering significant tidal disruption.  This suggests that, unlike
other clouds \citep[e.g., Orion, see][]{HB07}, TMC is not found in a
state of global collapse, explaining thus why it only forms low-mass
stars.  Small-scale collapse within the complex, on the other hand, is
permitted. Such local collapse enhances the formation of molecular gas
from H~I, and accounts for the rapid formation of stars
\citep{Heitsch_Hartmann08}.

Our result could be tested further if multi-epoch observations similar
to those obtained in Taurus by \citet{Loinard_etal07,
Torres_etal07,Torres_etal09} are performed for different star forming
regions exhibiting different efficiencies.  Such observations are
currently underway.

\section*{Acknowledgments}

We thank Paola D'Alessio and Lee Hartmann for a careful reading and
useful comments on this manuscript.  This work was supported by
UNAM-PAPIIT grant numbers 110606, IN111007, and IN119708 to JBP, LL,
and BP, respectively, and CONACYT grant numbers J50402-F, 47860, 50720
and 60581 to GCG,LL, BP and BP, respectively.  We have made extensive
use of the NASA-ADS database.  The calculations were performed both on
the cluster at CRyA-UNAM acquired with grant 36571-E, and on the
cluster Platform 4000 (KanBalam) at DGSCA, UNAM.

\label{lastpage}

 \end{document}